\documentclass[preprint,12pt]{elsarticle}

\usepackage{bm}
\usepackage{graphicx}

\usepackage{subcaption}

\usepackage{xfrac}

\usepackage{amssymb}

\usepackage{lineno}

\usepackage{placeins}

\newcounter{bla}

\journal{Computer Physics Communications}

\begin{document}
\begin{frontmatter}

\title{LBcuda: a high-performance CUDA port of LBsoft for simulation of colloidal systems}

\author[tov,iac]{Fabio Bonaccorso\corref{cor1}}
\ead{fabio.bonaccorso@roma2.infn.it }
\cortext[cor1]{Corresponding authors}

\author[iac]{Marco Lauricella\corref{cor1}}
\ead{marco.lauricella@cnr.it }

\author[tre,iac]{Andrea Montessori}

\author[cineca]{Giorgio Amati}

\author[iac]{Massimo Bernaschi}

\author[nvidia]{Filippo Spiga}

\author[iac]{Adriano Tiribocchi}

\author[iit,iac,harv]{Sauro Succi}

\address[tov]{Department of Physics and INFN, University of Rome Tor Vergata, Via della Ricerca Scientifica 1, 00133 Rome, Italy}

\address[iac]{IAC-CNR, Via dei Taurini 19, 00185 Rome, Italy}

\address[tre]{Dipartimento di Ingegneria, Università degli Studi Roma TRE, via Vito Volterra 62, Rome, 00146, Italy}

\address[cineca]{SCAI, SuperComputing Applications and Innovation Department, CINECA, Via dei Tizii, 6, Rome 00185, Italy}

\address[nvidia]{NVIDIA Development UK Ltd, Milton Hall, Ely Rd, Milton, Cambridge CB24 6WZ, United Kingdom}
\address[iit]{Center for Life Nano- \& Neuro-Science, Fondazione Istituto Italiano di Tecnologia (IIT), 00161 Rome, Italy}

\address[harv]{John A. Paulson School of Engineering and Applied Sciences, Harvard University, 33 Oxford St., Cambridge, MA 02138, USA}

\begin{abstract}
We present LBcuda, a GPU accelerated version of LBsoft, our open-source MPI-based software for the simulation of multi-component colloidal flows. 
We describe the design principles, the optimization and 
the resulting performance as compared to the CPU version, using both 
an average cost GPU and high-end NVidia GPU cards (V100 and the latest A100). The results show a substantial acceleration for the fluid solver reaching up to 200 GLUPS (Giga Lattice Updates Per Second) on a cluster made of 512 A100 NVIDIA cards simulating a grid of eight billion lattice points. 
These results open attractive prospects for the computational design 
of new materials based on colloidal particles.

\vspace{0.2cm}

\textbf{PROGRAM SUMMARY}
\vspace{0.2cm}

\begin{small}
\noindent
{\em Program Title:} LBcuda                                     \\
{\em CPC Library link to program files:} (to be added by Technical Editor)
\\
{\em Developer's repository link:} https://github.com/copmat/LBcuda
\\
\\
{\em Licensing provisions:} 3-Clause BSD License                                   \\
{\em Programming language:} CUDA Fortran              \\
{\em Nature of problem:} Hydro-dynamics of colloidal multi-component systems and Pickering emulsions.\\
{\em Solution method:} Lattice-Boltzmann method solving the Navier-Stokes equations for the fluid dynamics within an Eulerian description. Particle solver describing colloidal particles within a Lagrangian representation coupled to the fluid solver. The numerical solution of the coupling algorithm includes the back reaction effects for each force terms according to a fluid-particle multi-scale paradigm.
\\
\end{small}
\end{abstract}

\begin{keyword}
Lattice-Boltzmann \sep Colloids \sep CUDA \sep GPU

\end{keyword}

\end{frontmatter}


\section{Introduction}
\label{Introd}

In the last two decades, soft-glassy materials (SGM) have gained growing attention due to their applications in several industrial sectors. 
In particular, emulsions and foams are employed to design novel soft mesoscale materials for chemical, food processing, manufacturing, and biomedical purposes \cite{fernandez2016fluids,piazza2011soft,mezzenga2005understanding}.
Besides the technological relevance, their major significant theoretical interest stems from their intriguing non-equilibrium effects, 
including long-time relaxation, yield-stress behavior, and highly non-Newtonian dynamics. 

In this context, computational fluid dynamics (CFD) provides a valuable tool to improve the knowledge of the underlying physics of SGM.
To that purpose, a reliable SGM model alongside its software implementation is of apparent interest for the rational designing and shaping up of novel soft porous materials.

In this paper, we present and make available the CUDA Fortran code \textbf{LBcuda}, specifically designed to simulate on GPUs bi-continuous systems with colloidal particles under a variety of different conditions. \textbf{LBcuda} is a direct port of the LBsoft code \cite{lbsoft}, an open-source software for simulations of soft glassy emulsions originally developed for CPU-architectures, which successfully combines the lattice-Boltzmann method (LBM) \cite{succi2018lattice,kruger2017lattice,benzi1992lattice} with a Lagrangian solver to tackle the multi-scale coupling of fluids and particles \cite{bernaschi2019mesoscopic}.

Nowadays, the straightforward parallelization of LBM makes the lattice-Boltzmann algorithm an excellent candidate for high-performance CFD, especially on GPU-based architectures, given the relative simplicity and locality of its underlying algorithm.
As a consequence, several LBM implementations have been developed for GPU architectures, both academic packages 
such as the GPU-enabled versions of WaLBerla \cite{bauer2021walberla,holzer2021highly}, Palabos \cite{latt2021palabos}, Ludwig \cite{desplat2001ludwig}, MUPHY \cite{bernaschi2009muphy}, and commercially 
licensed software such as XFlow 2021 \cite{holman2012solution}, 
to name a few. 

\begin{figure}[h!]
\begin{center}
\includegraphics[width=0.8\linewidth]{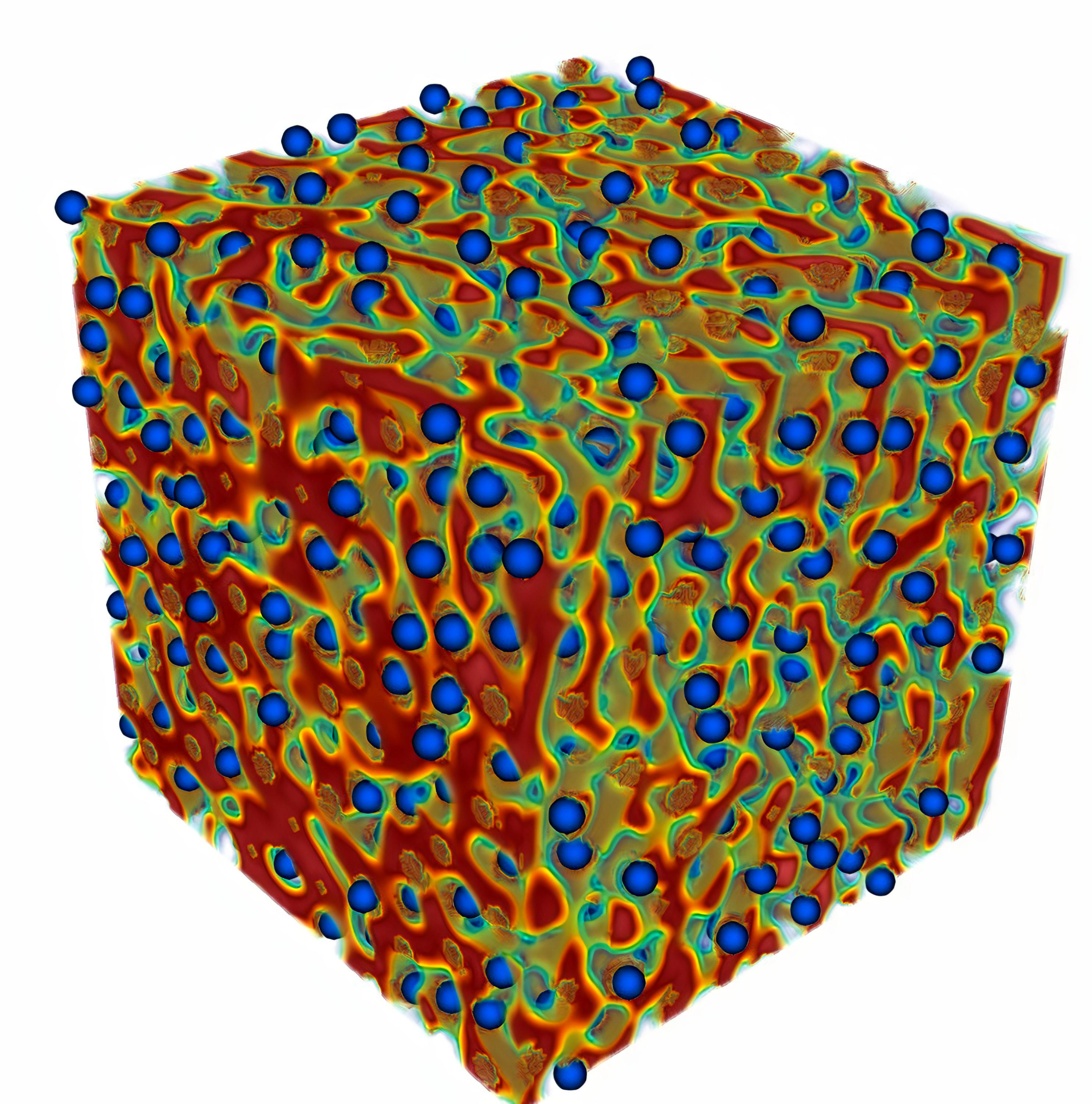}
\caption{A typical bijel configuration with colloidal particles (blue spheres) entrapping the interfaces between the two fluids (red and transparent green). }
\label{fig:bijel}
\end{center}
\end{figure}

As aforementioned, the model to describe colloidal particles is derived from the previous CPU-based LBSoft code, to which the reader is referred for further details \cite{lbsoft}. Briefly, SGM modeling requires specific implementations of LBM and Lagrangian solvers to include the hydrodynamic interactions between solid particles and fluids, following several strategies reported in the literature \cite{ladd2015lattice,ladd2001lattice,aidun1998direct,ladd1994numericala}. 
This extension has opened the possibility to simulate complex colloidal systems, also referred to as Pickering emulsions \cite{pickering1907cxcvi} which are of primary interest for the rational design of SGM \cite{xie2017direct,liu2016multiphase, frijters2012effects,jansen2011bijels}. This intrinsically multi-scale approach can catch, for example, the dynamical transition from a bi-continuous interfacially jammed emulsion gel, also referred to as bijel (see Fig. \ref{fig:bijel}), capturing the associated mechanical and spatial properties \cite{sun2021pickering}.

The paper is structured as follows. In Section \ref{sec:method} we report a very brief description of the underlying method, referring our previous paper for a deeper explanation of the details.
In Section \ref{sec:impl} we describe the details of the data structures on the GPU, while in Section \ref{sec:parallel} we explain the parallelization strategy and its impact on performance.
In Section \ref{perfomance} we report a set of tests used to validate the implementation and we investigate the performance against the reference CPU version (LBsoft). Finally, conclusion and outlook on future development directions are discussed.

\section{Method}
\label{sec:method}

In this section, we briefly review the approach to the simulation of SGM implemented in LBcuda alongside the more significant algorithmic adaptations required by the GPU-based hardware. A more detailed illustration of the underlying algorithms can be found in Ref. \cite{lbsoft}.  

The code combines two different levels of description: the first exploits a continuum approach for the dynamics of immiscible fluids, whereas the second manages individual rigid bodies representing colloidal particles or other suspended species. 
The two levels exchange information at each step of the time integration scheme to describe the concurrent interaction among particles and surrounding fluids.

In the first level, the LBM exploits a fully discretized analog of the Boltzmann kinetic equation to model flows and hydrodynamic interactions in fluids.

In the LBM approach, the fundamental quantity is $f_{i}(\vec{r};t)$, namely the probability of finding a “fluid particle” at the spatial point mesh $\vec{r}$ and at time $t$ with velocity  $\vec{c}_i$ selected from a 
finite set of possible speeds. The LBcuda code implements the 3-dimension 19-speed cubic lattice scheme (D3Q19) with the discrete velocities $\vec{c}_i$ with  $i \in [0,...,18]$ connecting mesh points with spacing $\Delta x$ (length lattice unit) to first and second mesh neighbours, located at distance $\Delta x$ and $\sqrt 2 \Delta x$, respectively (in other words, D3Q19 neglects 8 out of 27 possible velocities: those having distance $\sqrt 3\Delta x$).

Denoted $\rho(\vec{r};t)$ and $\vec{u}(\vec{r};t)$ respectively the fluid density and the fluid velocity, the lattice-Boltzmann equation is implemented in single-relaxation time (Bhatnagar-Gross-Krook equation) as follows:
\begin{equation}\label{eq:bgk}
f_{i}(\vec{r}+\vec{c}_{i};t+1)=(1-\omega)f_{i}(\vec{r};t)+\omega f_{i}^{eq}(\rho(\vec{r};t),\vec{u}(\vec{r};t))
\end{equation}
where $f^{eq}$ is the lattice local equilibrium, basically the local Maxwell-Boltzmann distribution (see Appendix A), and $\omega$ is a frequency tuning the relaxation towards the local equilibrium on a timescale $\tau=1/\omega$. The relaxation frequency $\omega$ controls the kinematic viscosity of the fluids according to the relation:
\begin{equation}
\nu=c_{s}^{2}\frac{\Delta x^{2}}{\Delta t} \left(\frac{1}{\omega}-\frac{1}{2}\right),
\label{eq:viscosity}
\end{equation}
where $\Delta x$ and $\Delta t$ are the physical length and time of the correspondent counterparts in lattice units. Note that the positivity of the kinematic viscosity requires the condition $0<\omega<2$.

In order to model a two component systems we adopted, a color gradient (CG) algorithm, which enforces a diffuse interface between the two fluids \cite{leclaire2017generalized}. 
In short, in the update phase of the populations, the CG collision contains three sub-steps: a plain BGK collision, a perturbation operator, and a final recoloring step. It is worth stressing that the last two sub-steps act only near the interface between the two fluids. Further details are reported in Appendix.

The second level of description involves a Lagrangian solver for the particle evolution, where each particle (colloid) is represented by a closed surface ${\mathcal S}$, taken, for simplicity, as a rigid sphere in the following. 

The LBcuda code adopts the formulation given by Jansen and Harting \cite{jansen2011bijels}, where only the exterior regions are filled with fluid, whereas the interior parts of the particles are solid nodes.
The solid--fluid interaction is managed via a simple generalization of the bounce-back rule including the correction due to the relative motion of the solid particle with respect to the surrounding fluid medium.

Hence, the particle position, speed $\vec{v}_p$ and angular momentum\index{angular!momentum} $\vec{\omega}_p$ are updated according to Newton's equations of motion:

\begin{equation}
\label{PARDYN}
\left\{ \begin{array}{lll}
\frac{d \vec{r}_p}{d t} =  \vec{v}_p,\\
m_p \frac{d \vec{v}_p}{d t} =  \vec{F}_p,\\
I_p \frac{d \vec{\omega}_p}{d t} = \vec{T}_p,
\end{array}   
\right.
\end{equation}
where $m_p$ and $I_p$ are the particle mass and moment of inertia, respectively.

Following Ladd's seminal works \cite{ladd1994numericala,ladd1994numericalb}, we advance in time eq. \ref{PARDYN} with a leap-frog scheme, which is second order accurate in time.
This set of equations considers the full many-body hydrodynamic interactions since the forces and torques are computed with the actual flow, as dictated by the presence of all $N$ particles simultaneously.

\section{Implementation}
\label{sec:impl}

The code is implemented in CUDA Fortran, using modules to minimize code cluttering. The LBcuda code requires no external libraries besides the CUDA runtime and compiles using a simple Makefile. The code is written for the nvfortran compiler, with the GPU kernels confined in {\em cuf} extension files, whereas the I/O part and the main are coded in standard FORTRAN files.

The code is composed of 6 files:
\begin{itemize}
\item \textbf{dimension.cuf}, which sets constants for the LB algorithm and the physical values of the simulation
\item \textbf{kernels\_fluid.cuf}, containing all GPU variables
\item \textbf{kernels\_fluid\_CG.cuf}, containing the color-gradient GPU code 
\item \textbf{kernels\_fluid\_PART.cuf}, containing the particles GPU code
\item \textbf{write\_output.f90}, which outputs the VTK and VTI files for external visualization by  graphical programs (e.g, ParaView) 
\item \textbf{main.f90}, finally contains the driving code of all the subroutines.
\end{itemize}

Most of the input for the simulation is defined by setting Fortran \textbf{parameters} in dimension.cuf. In contrast, other runtime parameters, such as print frequency of VTK output files and average statistical quantities, can be set up without recompilation in a plain-text input file.

The data for each fluid component are organized in a five dimension matrix having x,y,z, then the population index, and finally two possible values for \textit{switching} between old and new values during the collide-stream phases of the LB algorithm, also referred to as one-step two-grid algorithm \cite{wittmann2013comparison}.

All data residing on the GPU are defined in kernels\_fluid.cuf, whereas writing output files requires just a few memory passages from GPU to CPU at the printing frequency for fluid densities, flow field, and particle positions.

In order to solve the lattice-Boltzmann equation with particle dynamics, the algorithm proceeds executing the following sequence of subroutine calls:
\begin{itemize}
    \item Each thread of the GPU device computes fluid and particle quantities at a single spatial point, say located at the (i, j, k) node;
    \item In each node, the code proceeds according to three different cases:
    \begin{enumerate}
    \item If the node contains fluid far away from any particles, the thread will only advance the LB algorithm;
    \item If at that point a fluid touches a particle, the thread computes its part for the LB algorithm, then it computes its contribute for the force/toque integral. Note that when there are touching particles, a point can contribute to more than one particle.
    \item If the point is inside a particle, no computation is performed and the following steps are skipped;
    \end{enumerate}
    \item Apply the collision step of Eq. \ref{eq:bgk};
    \item Apply the halfway bounce-back rule at particle surface and the relative force terms on particles;
    \item Evolve position and angular velocity of particles (if present);
    \item Apply the stream step of Eq. \ref{eq:bgk}.
\end{itemize}
It is worth stressing that the net force and torque exerted from the fluid on the particle center of mass is obtained by summing over all the particle surface nodes.

\section{Parallelization strategy: CUDA and MPI}
\label{sec:parallel}

\begin{figure}
\begin{center}
\includegraphics[width=1.0\linewidth]{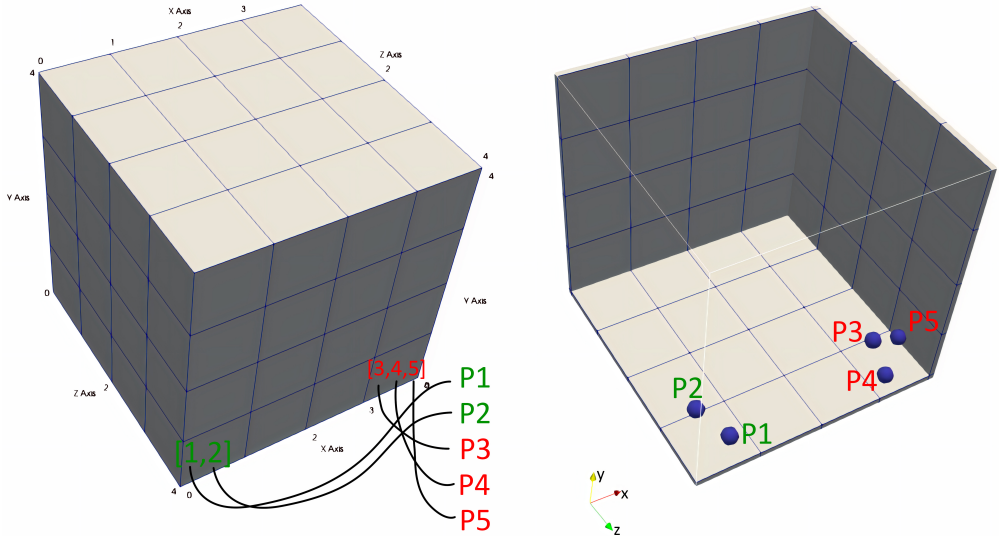}
\end{center}
\caption{Sketch showing the domain decomposition strategy used for the particle data on the GPU device. On the left, each sub-domain (thread block) has a list of owned particles. On the right, the global GPU data vector stores all the particles in contiguous way over the sub-domains.
}
\label{Figdec}
\end{figure}

For the LB part of the algorithm, the CUDA porting decomposes the global domain according to a 3D block distribution among the CUDA threads (see Fig. \ref{Figdec}). The selection of the block distribution is fixed at compile-time, and it can be tuned to obtain the best performance given the global grid dimension and the compute capability of the GPU device. Usually, a 3D decomposition that emphasizes the x-axis dimension achieves the best performance since it exploits the high memory bandwidth due to the data continuity in the column-major order of the FORTRAN language. Each thread will be responsible for only one grid point of the fluid box in each CUDA block. In the LB part, the thread iterates over all the fluid populations in most kernels. On the other hand, the thread computes the contribution of the fluid grid point to the force and torque of the overlapping particles, defined as particles whose surface overlaps the fluid node owned by the thread.


The LBcuda code is designed to exploit multiple GPU devices.
To that purpose, the code resorts to MPI having one GPU card associated to each MPI task. The LB domain is divided into sub-domains of equal size, whereas the variables related to the particles are replicated in all the MPI processes. Thus, the LB solver proceeds locally on each GPU device, with the extra computational cost due to the communication of border information among the local sub-domains of the neighbor MPI tasks. 

Each MPI task computes the part (section) of the particles falling in its sub-domain in the particle solver. 
In this framework, particles evolution is crucial for achieving a high computational throughput by avoiding excessive communication (or memory conflicts) among threads while integrating the particle quantities.
Thus, we have adopted a single particle list, which is stored on GPU devices. In particular, whenever the multi GPU is used, each MPI sub-domain has its list of owned particles on its GPU card. When a LB time step is completed, each thread in the sub-domain checks if it needs to compute the contribution of its fluid node to the computation of the surface integral for the force and torque that the fluid exerts on each particle surface node and vice versa.
Hence, a global MPI reduction is used to compute the corresponding total force and torque acting on the center of mass of each particle, so that particle positions, orientations, and velocities can be advanced in time on all the MPI processes. 
The selection of the sub-domain particle list on each MPI process is made by a CUDA kernel, leveraging the parallel computing power of each GPU card that makes the required computing time almost negligible.

It is worth highlighting that the overlap between particle and fluid node provides an unbalance in the work performed by the CUDA threads, notably wasting computational power at the surface particle node to treat the boundary conditions and the momentum exchange between the surrounding fluid and the particle. Nonetheless, we found that the overhead for the fluid solver is quite marginal, retaining an acceptable code scalability.

\section{Performance results}
\label{perfomance}

In order to analyze the computational performance of LBcuda, we consider two cubic boxes of side 128 and 256 lattice nodes
with periodic boundary along the three Cartesian axes. 
We limited the size to a 256 cubic box, which is the largest grid that fits in the memory of a single GPU device.
We perform three different test cases. First, we examine the case of a single fluid in a cubic box with an initial density equal to one and zero velocity flow ($case \; 1$).
As a second test ($case \; 2$), a bi-component system is considered where all the fluid nodes are randomly filled with fluid mass density of the two fluids, red and blue component, to achieve the value of the order parameter, $\phi=\frac{\rho_r-\rho_b}{\rho_r+\rho_b}$, equal to 1 or -1 with zero velocity flow field with the subscripts r and b standing for “red” and “blue” fluids, respectively. The third test ($case \; 3$) checks the entire LBcuda algorithm with
the two-component fluid combined with the particle solver 
describing the colloids in a rapid demixing emulsion.
In particular, we defined three sub-setups with different numbers of particles to assess the performance of the particle solver.
Hence, three values of the volume fraction occupied by particles are considered:
0.1\%, 1.0\% and 10\%, labeled $case \; 3a$, $case \; 3b$, and $case \; 3c$, respectively.

To evaluate the performance of LBcuda, we compare the theoretical peak performance to the actual one achieved by our code. In particular, the roofline model \cite{williams2009roofline} is used to rank the achievable computational performance in terms of Operational Intensity (OI), defined as the ratio between flops performed and data that need to be loaded/stored from/to memory. 
At low OI (say, $ < 10 $), the performance is limited by the memory bandwidth, whereas for higher OI values, the limitation comes from the availability of floating-point units. 
It is well known that LB is a bandwidth-limited numerical scheme, like most CFD models \cite{towards}. The OI index for LB schemes is around 0.7 for double-precision (DP) simulations using a D3Q19 lattice. As a matter of fact, for a single fluid, since the number of floating-point operations per lattice site and time step is $F \simeq 200 \div 250 $ and the load/store burden in bytes is $B= 19 \times 2 \times 8=304 $ (using double precision), the operational intensity is $F/B \sim 0.7$, whereas in single precision is $F/B \sim 1.4$ confirming that the code is bandwidth limited (see also Figure \ref{fig:roof}).

In the following, we assess the efficiency of the LBcuda code by means of the Giga Lattice Updates Per Second (GLUPS) metrics. In particular, 
the definition of GLUPS reads:
\begin{equation}
\label{eq:glups}
\text{GLUPS}=\frac{L_x L_y L_z}{10^9 t_{\text{s}}},
\label{eq:mlups}
\end{equation}
where $L_x$, $L_y$, and $L_z$ are the domain sizes in the
$x-$, $y-$, and $z-$ axis, and $t_{\text{s}}$ is the run (wall-clock) time (in seconds) per single time step iteration.

\subsection{Single fluid}

The $case \; 1$ with one fluid is tested using a plain BGK collision and a fused implementation (in which the collision and streaming step are performed simultaneously) of the LB time-advancing, the latter being the most popular approach for major LB codes. The results are reported for two different GPUs: a V100 and a GeForce RTX 2060S  in Table \ref{tab:1}.

\FloatBarrier
\begin{table}[h]
    \centering
    \begin{tabular}{|c|c|c|c|c|}
        \hline 
        \textbf{Size} & \textbf{GPU} & \textbf{Time} & \textbf{GLUPS} & \textbf{Approach} \\
        \hline 
        $128^3$ & V100  &  1.09 ms/iter & 1.923 & fused \\
        $128^3$ & 2060S &  1.64 ms/iter & 1.278 & fused \\
        \hline
        $256^3$ & V100  &  8.34 ms/iter & 2.011 & fused \\
        $256^3$ & 2060S &  15.6 ms/iter & 1.075 & fused \\
        \hline 
        $128^3$ & V100  &  2.09 ms/iter & 1.003 & plain \\
        $128^3$ & 2060S &  3.14 ms/iter & 0.667 & plain \\
        \hline
        $256^3$ & V100  &  15.34 ms/iter & 1.093 & plain \\
        $256^3$ & 2060S &  30.1 ms/iter & 0.557 & plain \\
        \hline 
    \end{tabular}
    \caption{Timings alongside with GLUPS of $128^3$ and $256^3$ cubic boxes for the $case \; 1$ using both the optimized fused and plain approach single fluid on a Tesla V100 and a GeForce RTX 2060S in single precision.}
    \label{tab:1}
\end{table}
\FloatBarrier

The key differences between the two GPUs are: the V100 has 5132 Cuda cores offering a peak performance of 14TFlop/s and a memory bandwidth of 900GB/s, whereas the GeForce RTX 2060S has 2176 cores for a peak of 6.4TFlop/s and a memory bandwidth of 448GB/s. On the other hand, we observe that the obtained 2.011 GLUPS for a cubic box of side 256 is comparable with the state-of-the-art represented, for instance, by the highly optimized code by G. Falcucci et al. in Ref. \cite{falcucci2021extreme} which reaches 3.406 GLUPS on a single V100 with the fused implementation on the same cubic box size.

Although the fused approach reduces of a factor two the number of memory accesses, it is worth highlighting that the particle solver requires the mandatory use of the plain approach, showing a decrease of the performance of about a factor two for $case \; 1$.
Indeed, the particle boundaries require using the plain Lattice-Boltzmann algorithm, where the collision is first computed for all the fluid lattice nodes. Then, the boundary conditions are applied (internal walls or particles), and finally the streaming of the populations is carried out (see Section \ref{sec:impl}).

\begin{figure} \begin{center}
\includegraphics[width=1.0\linewidth]{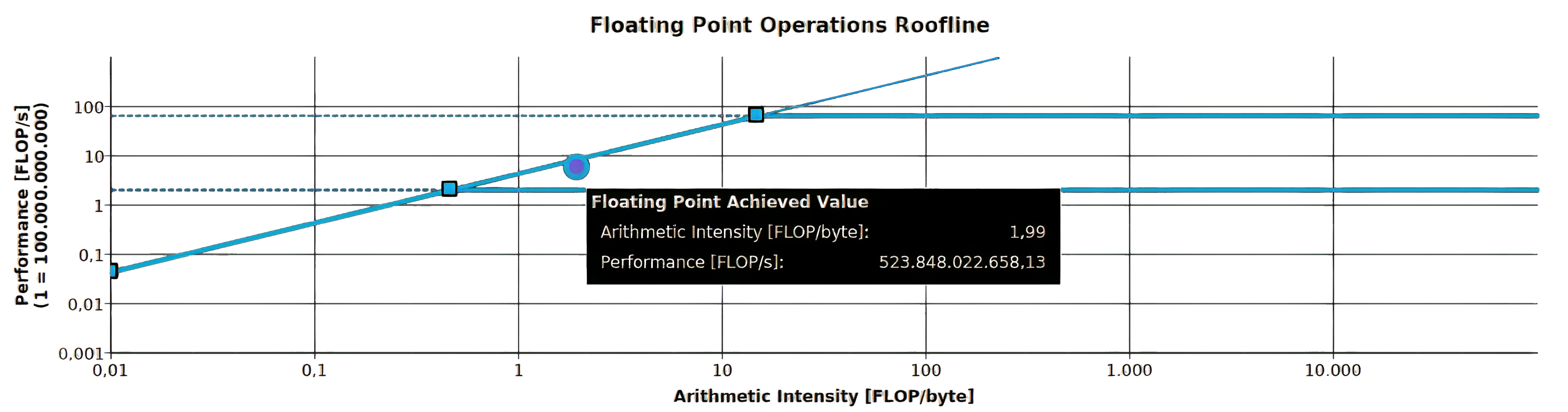}
\end{center}
\caption{Roofline model for single fluid using a GeForce RTX 2060S, as measured by NVIDIA NSight in the plain LB approach and in single precision. Bandwidth and Float point computation limits were obtained performing {\em memory-stream} and High Performance Computing Linpack benchmark. Note that the LBcuda code lies on the left part of the plot showing that it is bandwidth limited.}
\label{fig:roof}
\end{figure}

From an in-depth analysis using the NVidia dedicated tool (NSight), we observe that on the GeForce RTX 2060S the main kernel (the LB time-stepping before the streaming substep)128x1x1 achieves better performance  achieves almost $\sim 520 GFlop/s$ with an arithmetic intensity of 2.0 in single precision (see Figure \ref{fig:roof}). This result shows that we are far from an intensity $\sim 15.0$ which should give the peak performance, and in the $case \; 1$ we attain about 60\% of memory bandwidth utilization due to its non-optimal use following the plain approach.

\subsection{Two fluid test}

The $case \; 2$ is related to the simulation of a two-component system by the color-gradient model (see Section \ref{sec:method}). We remark that the time integration is implemented using the plain approach with a standard collide-stream 2-pass algorithm. Table \ref{tab:cg} highlights the measured performance for the $128^3$ and $256^3$ cubic box. In particular, we observe an increase of about a factor 3 with respect to the previous $case \; 1$, which is mainly due to the larger number of operations, more than doubled, in the color gradient collision operator containing three steps (see Section \ref{sec:method}) instead of the single step of the plain BGK single fluid case. 

\FloatBarrier
\begin{table}[h]
    \centering
    \begin{tabular}{|c|c|c|c|c|}
        \hline 
        \textbf{Size} & \textbf{GPU} & \textbf{CUDA Block} & \textbf{Time} & \textbf{GLUPS} \\
        \hline 
        $128^3$ & V100  & 8x4x4 &  4.5 ms/iter & 0.466 \\
        $128^3$ & 2060S & 8x4x4 &  11 ms/iter & 0.190 \\
        \hline
        $256^3$ & V100  & 8x4x4 &  34.1 ms/iter & 0.492 \\
        $256^3$ & 2060S & 8x4x4 &  82.3 ms/iter & 0.204 \\
        \hline 
        \hline 
        $128^3$ & V100  & 128x1x1 &  3.96 ms/iter & 0.554 \\
        $128^3$ & 2060S & 128x1x1 &  8.96 ms/iter & 0.245 \\
        \hline
        $256^3$ & V100  & 128x1x1 &  27.9 ms/iter & 0.601 \\
        $256^3$ & 2060S & 128x1x1 &  64.5 ms/iter & 0.261 \\
        \hline 
    \end{tabular}
    \caption{Timings alongside with GLUPS for $128^3$ and $256^3$, 2 fluids with CG using a Tesla V100 and a GeForce RTX 2060S with the plain approach in single precision with two different decompositions of threads in CUDA block. Note that the CUDA block configuration 128x1x1 achieves better performance exploiting the contiguous data over $x$ in the population arrays.}
    \label{tab:cg}
\end{table}
\FloatBarrier

\subsection{Particles}

The entire LBcuda algorithm is evaluated with the two-component colour gradient method (see Section \ref{sec:method}) combined
with the particle solver to model the colloids in a rapid demixing emulsion. To assess
the performance of the particle solver, we prepared three simulation setups with different numbers of particles with radius equal to 5.5 lattice units in a cubic box of side 256 lattice points.
The three cases, in the following labelled $case \; 3a$, $3b$, and $3c$, differ in the ratio between the volume occupied by the particles compared to the box volume equal to $\varphi=$ 0.1\%, 1.0\% and 10\%, respectively, in order to study the impact of the particle evolution on the simulation time.
The performance impact of having particles goes from negligible in $case \; 3a$ to be comparable with the LB computation in $ case \; 3c$ in which the time for each iteration almost doubles, as shown in table \ref{tab:partic}.

\FloatBarrier
\begin{table}[h]
    \centering
    \begin{tabular}{|c|c|c|c|c|}
        \hline 
        \textit{\textbf{case}} & \textbf{GPU}  & $\varphi$ & \textbf{Time} & \textbf{GLUPS} \\
        \hline 
        $3a$ & V100  &  0.1 \% & 34.6 ms/iter &  0.484 \\
        $3a$ & 2060S &   0.1 \% & 84 ms/iter &  0.199  \\
        \hline 
        $3b$ & V100  &   1.0 \% & 37.6 ms/iter &  0.446 \\
        $3b$ & 2060S &   1.0 \% & 88 ms/iter &  0.190 \\
        \hline 
        $3c$ & V100  &   10 \% & 56 ms/iter &  0.299 \\
        $3c$ & 2060S &   10 \% & 113 ms/iter &  0.148 \\
        \hline
    \end{tabular}
    \caption{Timings alongside with GLUPS for the three cases with particle volume fraction, $\varphi$, equal to 0.1\%, 1.0\%, and 10\%, respectively, in a cubic box of $256^3$ lattice nodes in single precision.}
    \label{tab:partic}
\end{table}
\FloatBarrier

\begin{figure}[h!]
\includegraphics[width=1.0\linewidth]{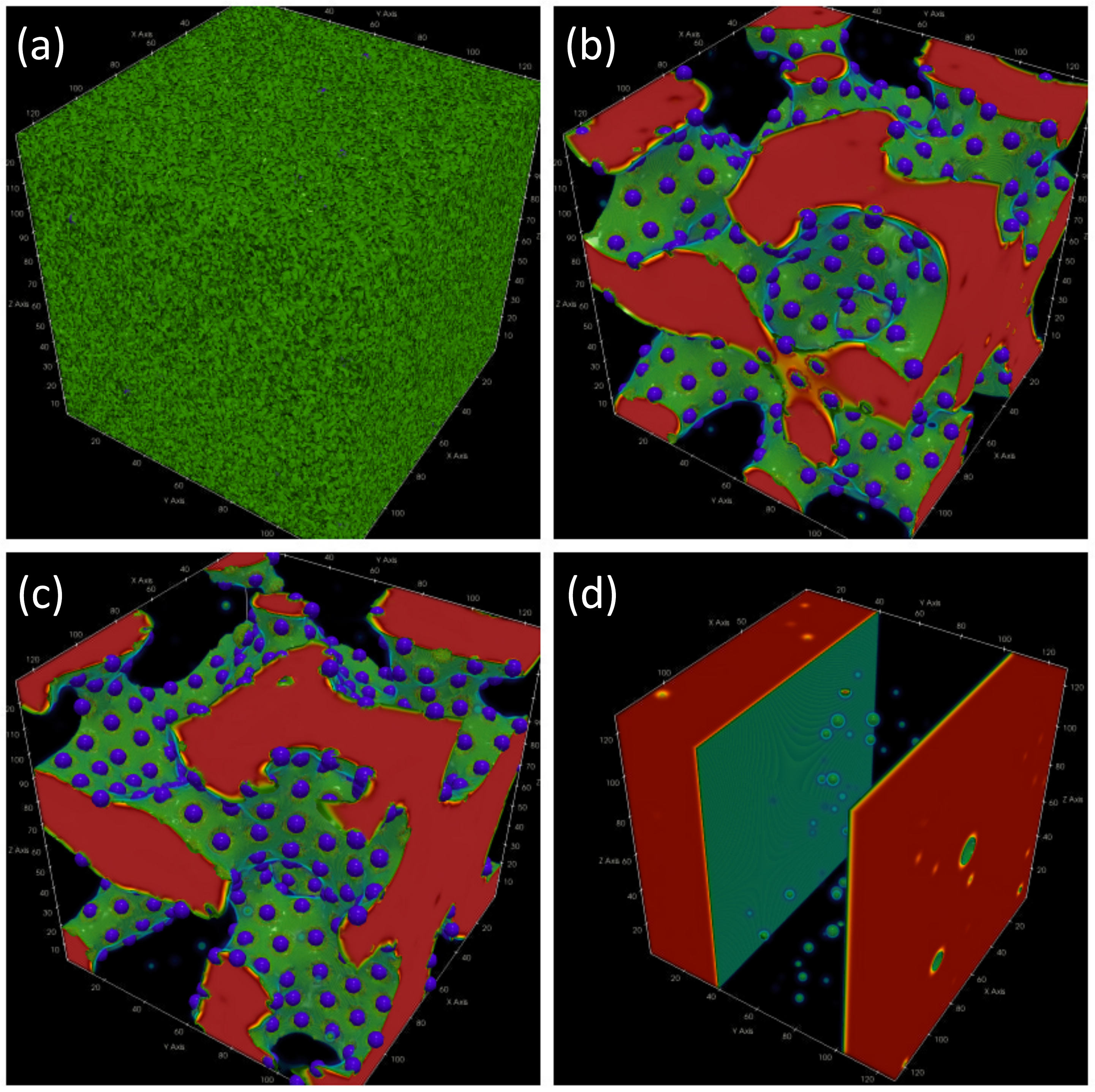}
\caption{Renderings for $256^3$ simulation without (plot  d) and with particles (plot a,b,c). From top-left to bottom-right) a) Initial condition for particle simulation.  b) Density field after 50k iterations with 10\% volume fraction.  c) Density field after 100k iterations (10\% vol. fraction).  d) Density field after 100k iterations without particles starting from a mixed bi-component fluid system (similar to plot a).}
\label{fig:stopInterface}
\end{figure}

It is worth highlighting that LBcuda code implements a double precision accumulator because of floating point accuracy problems related to the momentum transfer from the fluid to each particle. In particular, the particle force and torque computation suffer from floating accuracy problems due to strong cancellation between addends of alternating sign over the nodes of the two-fluids interface.

The benchmark results can be also compared to the corresponding box size without particles reported in Table \ref{tab:cg}.
For the test in $case \; 3$ the particle radius is equal to 5.5 lattice units, and
the particle positions are randomly distributed in the box. 
The initial particle velocity is zero in all runs.
The particle wettability is tuned to set an angle equal to $90^{\circ}$ 
with respect to the axis $\vec{x}^{\star}$ in the local reference frame 
of each particle. The impenetrability among particles is avoided by an hertzian repulsive contact force computed by means of neighbor's lists (see Section \ref{sec:parallel}).
The lubrication force is also considered by adding an extra force term whenever two particles are located at a mutual distance lower than 2/3 lattice unit, as reported in previous simulations \cite{jansen2011bijels}. 
The particle mass was estimated as the weight corresponding to a particle 
made of silica \cite{herzig2007bicontinuous}.

All runs were simulated on both the CPU and GPU architectures using the previous LBSoft code and the corresponding GPU ported version, LBcuda. In all the $cases \; 3$ we observe only a small deviation in the position always lower than $10^{-4}$ in lattice units, mainly due to the aforementioned floating point accuracy problem. Indeed, the order of the addends over the particle surface nodes is completely random on the GPU device. 
In the $case \; 3c$ we observe the arrest of the phase demixing process with particles located at the fluid-fluid interface entrapping the demixing process into a metastable state, the bi-continuous jammed gel state \cite{stratford2005colloidal}. In Figure \ref{fig:stopInterface} we show two of these numerical experiments, in which a random mixture of two fluids evolves in a very different way in the presence of a high number of particles ($case \; 3c$) and without particles ($case \; 2$). Indeed, the particles stop the complete spinodal decomposition of the two fluids corresponding to the condition of minimal energy and minimal interface between them. Whenever a high volume fraction of particles is in the simulation box, the separation surface becomes way more "corrugated" showing the formation of the bi-continuous jammed gel state.
Figure \ref{fig:stopInterface} shows the initial condition, the fluids after 50k (with particles), and final configurations of both simulations after 100k iterations ($case \; 3c$ and $case \; 2$).

\subsection{Multi GPUs Performance}

The LBcuda code resorts to the Message Passing Interface (MPI) library to exchange data among GPU devices running in parallel.
The performance obtained using multiple GPUs show a good scaling behavior as reported in Tables \ref{tab:multiGPU} and  \ref{tab:multiGPU2}. The benchmark in Table \ref{tab:multiGPU} was carried out on Marconi100 at CINECA, a cluster of V100 cards, each with 16 GB of global memory using a different number of GPU devices, for three cubic boxes: $256^3$, $512^3$, and $1024^3$. The cluster is made of nodes, each endowed with four GPU cards so that the MPI communication does not incur network latency unless the job is using more than 4 GPU cards. The benchmark in Table \ref{tab:multiGPU2} ran on a cluster made of NVIDIA DGX A100. Each NVIDIA DGX A100 is equipped with 8 A100 NVIDIA GPU with 80 GB of RAM interconnected intra-node through the NVswitch and 8 NVIDIA Infiniband HDR (one NIC for each GPU) for multi-node scaling.

\begin{table}[h!]
    \centering
    \begin{tabular}{|c|c|c|c|c|c|c|}
        \hline 
        \textbf{Grid} & \textbf{1} & \textbf{2} & \textbf{4}  & \textbf{8}  & \textbf{16} & \textbf{32} \\
        \hline 
     
        $256^3 V100\ 16GB$ & 0.60 & 1.17 & 2.24 & 2.93 & 2.71 & 3.15 \\
        \hline 
        $512^3 V100\ 16GB$ & x & x & 2.67 & 4.57 & 6.23 & 9.26 \\
        \hline
        $1024^3 V100\ 16GB$ &  x     &   x    &  x    &  x    &   x    & 12.48 \\
        \hline
        \end{tabular}
    \caption{Performance for $256^3$, $512^3$, $1024^3$, measured in GLUPS running on multiple NVIDIA Volta V100 GPUs (each card equipped with 16 GB of RAM). Note that the symbol (x) denotes a grid size too big to fit in the local GPU memory. For the $256^3$, performance degrade dramatically when using more than 8 V100 GPUs and more than 64 A100 because each local domain becomes too small (256x256x8) to offset the cost of scheduling the GPU task.}
    \label{tab:multiGPU}
\end{table}

\begin{table}[h!]
    \centering
      \begin{tabular}{|c|c|c|c|c|c|c|c|}
        \hline
        \textbf{Grid} & \textbf{8} & \textbf{16} & \textbf{32}  & \textbf{64} & \textbf{128} &  \textbf{256} & \textbf{512}   \\
        \hline
        $256^3 A100\ 80GB$ & 6.17 & 9.42 & 12.38 & 15.38 &  & & \\
        \hline 
        $512^3 A100\ 80GB$ & 7.63 & 14.13 & 23.13 & 36.05  & & &\\
        \hline
        $1024^3 A100\ 80GB$ & 11.65 & 20.57 & 28.59 & 49.11  & 76.91 & & \\
        \hline
        $2048^3 A100\ 80GB$ & x & x & x & 55.22  & 102.3 & 160.9 & 204.5 \\
        \hline
                
    \end{tabular}
    \caption{Performance for $256^3$, $512^3$, $1024^3$, $2048^3$ measured in GLUPS running on multiple NVIDIA Volta A100 GPUs (each card equipped with 80 GByte of RAM). Note that the symbol (x) denotes a grid size too big to fit in the local GPU memory.}
    \label{tab:multiGPU2}
\end{table}

\begin{figure}[h!]
\begin{center}
\includegraphics[width=0.5\linewidth]{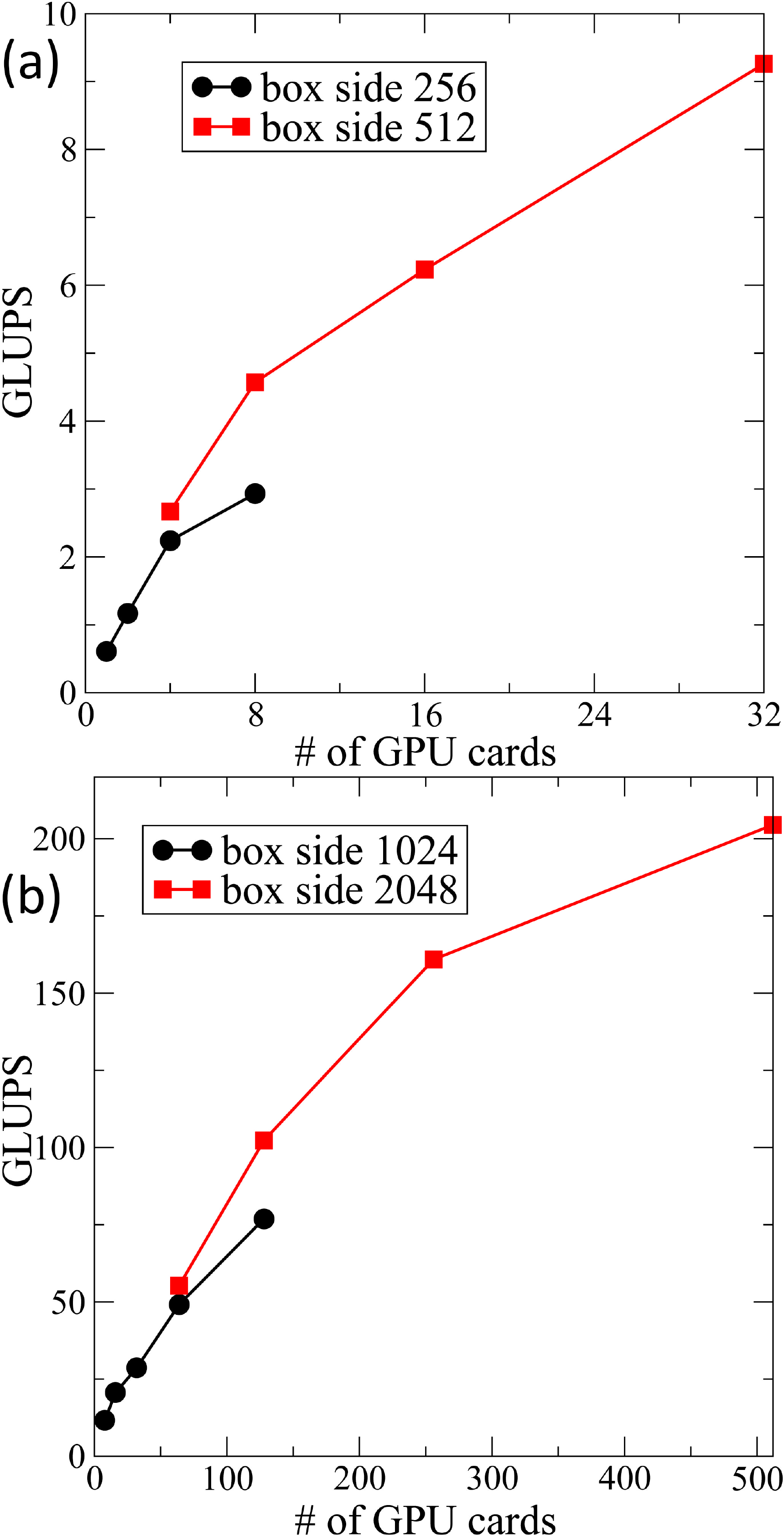}
\caption{Top panel: Measured GLUPS for different numbers of V100 cards with a 16 GB RAM, running two cubic boxes of side 256 and 512, respectively. Bottom panel: Measured GLUPS for different numbers of A100 cards with a 80 GB RAM, running two cubic boxes of side 1024 and 2048, respectively.}
\label{fig:GPUMlups}
\end{center}
\end{figure}

All benchmarks reported in Tables \ref{tab:multiGPU} and  \ref{tab:multiGPU2} were carried out on a bi-component fluid system without particles ($case \; 2$). 
The data for the cubic box sizes $256^3$, $512^3$ are also plotted in Figure \ref{fig:GPUMlups} with an evident decrease in the scalability whenever the code starts to run on more than one node (more than 4 GPUs).
Hence, the three tests, $case \; 3a$, $3b$, and $3c$, with different values in the particle volume fraction, $\varphi$, were performed to probe the efficiency of the particles solver as a function of the particles number in the system. All benchmarks were carried out on eight GPU cards: both V100 cards with a 16 GB RAM and A100 cards with a 40 GB RAM. In Table \ref{tab:partmultiGPU}, we observe a clear communication penalty as the particles number increases that is mainly due to the replicated data strategy used for the particle solver parallelization. Indeed, the replicated data parallel approach replicates the physical quantities of all particles across the MPI processes, performing local updating and global MPI sum reductions in order to advance the system in time, which decreases the performance as the quantity of particles data increases.
Nonetheless, the analysis of the performance as a function of the number of particles shows that the code is able to reach about 3.80 GLUPS in a system with 10\% in the particle volume fraction.

\begin{table}[h!]
    \centering
    \begin{tabular}{|c|c|c|c|c|}
        \hline 
        \textbf{GPU} & \textbf{No particles} & \bm{$\varphi=0.1\%$} & \bm{$\varphi=1.0\%$}& \bm{$\varphi=10\%$} \\
        \hline 
         8 V100@16 & 30 ms (4.47) & 40 ms (3.35) & 51 ms (2.63) & 95 ms (1.41) \\
        \hline
         8 A100@40 & 18 ms (7.55) & 19 ms (7.06) & 22 ms (6.10) & 35 ms (3.83) \\
        \hline
    \end{tabular}
    \caption{Time per single iteration alongside with GLUPS in parenthesis  for $512^3$ without and with particles (at different particle volume fraction $\varphi$)  on 2 different machines: using eight V100 with 16 GB of RAM on 2 nodes (connected by InfiniBand) and using eight A100 with 40 GB in a single node (without the latency time due to the InfiniBand communication).}
    \label{tab:partmultiGPU}
\end{table}

\subsection{Comparing LBcuda with LBsoft}

For the sake of completeness, we probe the gain provided by the CUDA port reported in the present article. Thus, the $case \; 2$ bi-component system was initialized with the same values in density and flow field in both LBcuda and LBsoft code. The cubic box size is equal to $512^3$ lattice points. Although it is difficult to compare two completely different computing architectures, we measure the wall-clock time obtained on a GPU cluster made of V100 cards with 16 GB RAM and a CPU cluster containing two Intel Cascade Lake 8260 CPUs at 2.40 GHz with 48 cores and 384 GB of RAM per node.
We note that the code produces the same results in single precision unless a slight difference of $10^{-4}$ order of magnitude in the particle positions due to floating point accuracy problems (mainly due to different order of summation). The wall-clock time for iteration results in 12 ms on 32 V100 cards of LBcuda versus 88 ms on 528 cores of LBSoft, confirming the clear advantage in running the CUDA ersion on a GPU HPC cluster.


\section{Conclusion}

We have presented LBcuda, a CUDA port of LBsoft, an open-source software aimed at simulating specifically colloidal systems. 
LBcuda is written in CUDA Fortran and
permits to simulate large system sizes running on multiple GPU devices by exploiting an efficient parallel domain decomposition implementation.

In particular, the code shows good scaling behavior of the fluid solver achieving the performance peak of 200 GLUPS on 512 NVIDIA A100 cards with a grid of eight billion lattice points.
On the other hand, the particle solver combined with the LB approach shows a very satisfactory performance in terms of scalability in both system size and number of processing cores, especially using the Nvidia Ampere A100 cards.

In this work, the main structure of LBcuda has been outlined along with the key steps of its implementation. 
Furthermore, several cases have been introduced to test the code over typical problems that the LBcuda code can deal with. 
In particular, the simulations with particles demonstrate the capabilities of the present code to reproduce the complex dynamics of bi-jel systems in a rapid de-mixing emulsion.

The LBcuda code is open source and completely accessible at the public repository GitHub, which is in line with the spirit of open-source software, mainly to promote the contribution of independent developers.

\section{Acknowledgments}
The research leading to these results has received
funding from the European Research Council under the European
Union's Horizon 2020 Framework Programme (No. FP/2014-
2020)/ERC Grant Agreement No. 739964 ``COPMAT'' and from MIUR under the project “3D-Phys” (No. PRIN 2017PHRM8X). The CINECA is acknowledged for the support granted by the ISCRA project ``porting LBSOft in CUda on multi-node GPUs (LBSOCU)''.

\section*{Appendix}

For the details of the color gradient (GC) model of the Lattice Boltzmann method employed in the bi-component systems \cite{leclaire2017generalized}, we recall some notions.

In the color gradient LB for two-component flows, two
sets of distribution functions are defined to track the evolution of the two fluid components, which occurs via a streaming-collision algorithm:

\begin{equation} \label{CGLBE}
f_{i}^{k} \left(\vec{x}+\vec{c}_{i}\Delta t,\,t+\Delta t\right) =f_{i}^{k}\left(\vec{x},\,t\right)+\Omega_{i}^{k}( f_{i}^{k}\left(\vec{x},\,t\right)),
\end{equation}

where $f_{i}^{k}$ is the discrete distribution function, representing
the probability of finding a particle of the $k^{th}$ component at position $\vec{x}$ and time
$t$ with discrete velocity $\vec{c}_{i}$ . 

In the last Eq. $i$ is the index running over the lattice discrete directions $i = 0,...,b$, where $b=18$ for a three dimensional 19 speed lattice (D3Q19) implemented in LBcuda.
The lattice time step $\Delta t$ has been taken as $1$ (in lattice units) for convenience.
The density $\rho^{k}$ of the $k^{th}$ component is given  by the zeroth moment of the distribution functions:
\begin{equation}
\rho^{k}\left(\vec{x},\,t\right) = \sum_i f_{i}^{k}\left(\vec{x},\,t\right),
\end{equation}
while the total fluid density is assessed as $\rho=\sum_k \rho^k$, and the total momentum of the mixture is given as the sum of the linear momentum of the two components:
\begin{equation}
\rho \vec{u} = \sum_k  \sum_i f_{i}^{k}\left(\vec{x},\,t\right) \vec{c}_{i}.
\end{equation}

The collision operator in the CG model is made of three parts: 

\begin{equation}
\Omega_{i}^{k} = \left(\Omega_{i}^{k}\right)^{(3)}\left[\left(\Omega_{i}^{k}\right)^{(1)}+\left(\Omega_{i}^{k}\right)^{(2)}\right].
\end{equation}

In the above, $\left(\Omega_{i}^{k}\right)^{(1)}$ stands for the standard collisional relaxation which reads:
\begin{equation}
\left(\Omega_{i}^{k}\right)^{(1)}=\omega(f_i^{k,eq} - f_i^k),
\label{1coll}
\end{equation}
where $\omega=2/(6\bar{\nu} -1)$ is the effective relaxation parameter being $\bar{\nu}$ the mean viscosity of the bi-component system computed as
$\frac{1}{\bar{\nu}}=\frac{\rho_1}{(\rho_1+\rho_2)}\frac{1}{\nu_1} + \frac{\rho_2}{(\rho_1+\rho_2)}\frac{1}{\nu_2}$ ($\nu_1$ and $\nu_2$ are the kinematic viscosities of the two pure components in the bulk). 
The equilibrium distribution function of the $k^{th}$ component $f_i^{k,eq}$ is given by a low-Mach, second-order, expansion
of a local Maxwellian, namely:
\begin{equation}
f_i^{k,eq}=w_i \rho^k (1 + \frac{ \vec{c_i} \cdot \vec{u}}{c_s^2} +\frac{(\vec{c_i} \cdot \vec{u})^2}{2c_s^4} - \frac{\vec{u} \cdot \vec{u}}{2 c_s^2}),
\label{eq:LEQ}
\end{equation}
where $c_s=1/\sqrt{3}$ is the sound speed of the model.
The symbol $\left(\Omega_{i}^{k}\right)^{(2)}$ denotes the perturbation step, which 
contributes to the build up of an interfacial tension. Finally, $\left(\Omega_{i}^{k}\right)^{(3)}$ is the recoloring step, which promotes the segregation  between species, so as to minimize their mutual diffusion.

In order to reproduce the correct form of the stress tensor, the perturbation operator can be constructed by exploiting the concept of the continuum surface force.
Firstly, the perturbation operator must satisfy the following  conservation constraints:

\begin{eqnarray} \label{consconstr}
\sum_i \left(\Omega_{i}^{k}\right)^{(2)}=0 \\
\sum_k \sum_i \left(\Omega_{i}^{k}\right)^{(2)} \vec{c}_i=0
\end{eqnarray}

By performing a Chapman-Enskog expansion, it can be shown that the hydrodynamic limit of Eq.\ref{CGLBE} is represented by a
set of equations for the conservation of mass and linear momentum:

\begin{eqnarray} \label{NSE}
\frac{\partial \rho}{\partial t} + \nabla \cdot {\rho \vec{u}}=0 \\
\frac{\partial \rho \vec{u}}{\partial t} + \nabla \cdot {\rho \vec{u}\vec{u}}=-\nabla p + \nabla \cdot [\rho \nu (\nabla \vec{u} + \nabla \vec{u}^T)] + \nabla \cdot \bm{\Sigma}
\end{eqnarray}

where $p=\sum_k p_k$ is the pressure and $\nu=c_s^2(\tau-1/2)$ is the kinematic viscosity of the mixture, being $\tau$ the single relaxation time.

The stress tensor in the momentum equation is given by:

\begin{equation}
\bm{\Sigma}=-\tau \sum_i \sum_k \left( \Omega_{i}^{k} \right)^{(2)} \vec{c}_i \vec{c}_i
\end{equation}

Since the perturbation operator is responsible for generating interfacial tension, the following 
relation must hold:

\begin{equation} 
\nabla \cdot \bm{\Sigma} = \vec{F}
\label{SeqF}
\end{equation}

Denoting $\Theta =(\rho^1-\rho^2)/(\rho^1+\rho^2)$ 
the phase field, by choosing the second operator as:
\begin{equation} 
\left(\Omega_{i}^{k}\right)^{(2)}= \frac{A_k}{2} |\nabla \Theta|\left[w_i \frac{(\vec{c}_i \cdot \nabla \Theta)^2}{|\nabla \Theta|^2} -B_i \right], 
\label{2coll}
\end{equation}
substituting it into Eqs \ref{consconstr} and \ref{SeqF} and by imposing that the set $B_i$ must satisfy the following isotropy constraints

\begin{eqnarray}
\sum_i B_i= \frac{1}{3} \; ; \sum_i B_i \vec{c}_i=0 \;  ; \sum_i B_i \vec{c}_i \vec{c}_i= \frac{1}{3} \mathbf{I},
\end{eqnarray}

we obtain an equation for the surface tension of the model:
\begin{equation}\label{sigmaA}
\sigma=\frac{2}{9}(A_1+A_2)\frac{1}{\omega}=\frac{4}{9}A\frac{1}{\omega}.
\end{equation}
The above relation shows a direct link between the surface tension and the parameter $A$ with $A_1=A_2$.
In actual practice, after choosing the viscosity of the two components and the surface tension of the model, at each time step, one locally computes the $A$  coefficient by using the formula reported in eq. (\ref{sigmaA}).

As pointed out above, the perturbation operator generates an interfacial tension in compliance with the capillary-stress tensor of the Navier-Stokes equations for a multicomponent fluid system.

Nonetheless, the perturbation operator alone does not guarantee the immiscibility of different fluid components.
For this reason, a further step is needed (i.e. the recoloring step) to minimize the mutual diffusion between components.

Following the work of Latva-Kokko and
Rothman, the recoloring operator for the two sets of distributions takes the following form:

\begin{eqnarray}
\left(\Omega_{i}^{1}\right)^{(3)} =\frac{\rho^1}{\rho} f_i^* + \beta \frac{\rho^1\rho^2}{\rho^2} \cos{\phi_i} f_i^{eq,0} \\
\left(\Omega_{i}^{2}\right)^{(3)} =\frac{\rho^2}{\rho} f_i^* - \beta \frac{\rho^1\rho^2}{\rho^2} \cos{\phi_i} f_i^{eq,0},
\label{3coll}
\end{eqnarray}

where $f_i^*=\sum_k f_i^{k,*}$ denotes the set of post-perturbation distributions, $\rho=\rho^1 + \rho^2$,  $\cos{\phi_i}$ is the angle between the phase field gradient and the $i^{th}$ lattice vector and $f_i^{eq,0}=f_i(\rho,\vec{u}=0)^{eq}=\sum_k f_i^k(\rho,\vec{u}=0)^{eq}$ is the total zero-velocity equilibrium distribution function. In Eq. \ref{3coll}, the coefficient $\beta$ is a free parameter which tunes the interface width, thus playing the role of an inverse diffusion length scale. The coefficients used in Eqs \ref{eq:LEQ} and \ref{2coll} are reported in Table.

\begin{table}
    \centering
    \begin{tabular}{ c c c c }
        \hline 
        \hline 
        D3Q19 & $\{ i:\left| c \right|^2 = 0 \}$ & $\{ i:\left| c \right|^2 = 1 \}$ & $\{ i:\left| c \right|^2 = 2 \}$ \\
        \hline 
        \hline 
         $w_i$ & 1/3 &  1/18 & 1/36  \\
        
         $B_i$ & -2/9 & 1/54 & 1/27  \\
        
        \hline
    \end{tabular}
    \caption*{D3Q19 lattice velocity and weights \cite{leclaire2017generalized}.}
    \label{tab:coef}
\end{table}

For the details of the particle time evolution, a full explanation is in Ref. \citep{lbsoft}. For the sake of completeness, the main key steps are outlined in the following.

The velocity at a boundary node of $p$-th particle is given by:
\begin{equation}
\label{UPI}
\vec{u}_{b} = \vec{v}_p + \vec{r}_b \times \vec{\omega}_p,
\end{equation}
where, $r_{b} = \vec{r}_s + \frac {1}{2} \vec{c}_i$ is the 
location of the moving wall along the $i$-th link connecting
the solid node $\vec{r}_s$ to the 
fluid node $\vec{r}_i = \vec{r}_s +\vec{c}_i$.
All coordinates are relative to the center of the $p$-th particle, located at position $\vec{r}_p$ and moving with translation and angular velocities $\vec{v}_p$ and $\vec{\omega}_p$, respectively.

The timestep is made unit for simplicity. 
This velocity sets the bias between colliding pairs:
\begin{eqnarray}
\label{LADDFLBE}
f_i \left( \vec{r}+\vec{c}_i, t+1 \right) =
f_{\bar i}  \left( \vec{r}+\vec{c}_i, t' \right) + 2 \rho w_i u_{bi}, \\
f_{\bar i} \left( \vec{r}, t+1 \right) =  f_i \left( \vec{r}, t' \right) 
- 2 \rho w_i u_{bi}
\end{eqnarray}
where $t'$ denotes post-collisional states and we have set
$$
u_{bi}= \frac{\vec{u}_b \cdot \vec{c}_i}{c_s^2}
$$
Note that these rules reduce to the usual bounce-back conditions for a solid at rest, $\vec{v}_p=\vec{\omega}_p=0$.

These collision rules produce a net momentum transfer between the fluid and the solid site:
\begin{equation}
\label{FORCEPP}
\vec{F}_i(\vec{r}_b,t+\frac{1}{2}) = 2 \vec{c}_i 
[f_i (\vec{r},t') - f_{\bar i}(\vec{r}_i,t') - 2 \rho w_i u_{bi}]
\end{equation}

The net force acting upon particle $p$ is obtained by summing over all boundary sites $\vec{r}_b$ and associated interacting links, namely:
\begin{equation}
\label{FORCEPARTEQ}
\vec{F}_p = \sum_{{\vec{r}_b,i} \in \Sigma_p} \vec{F}_{i}(\vec{r}_b).
\end{equation}
Similarly, the total torque\index{torque} $\vec{T}_p$ is computed as
\begin{equation}
\label{TORQUEPARTEQ}
\vec{T}_p = \sum_{{\vec{r}_b,i} \in \Sigma_p} \vec{F}_{i}(\vec{r}_b)
\times \vec{r}_b,
\end{equation}


\end{document}